# Parallel execution of portfolio optimization

R. Nuriyev

## Abstract

*Analysis of asset liability management (ALM) strategies especially for long term horizon is a crucial issue for banks, funds and insurance companies.*

*Modern economic models, investment strategies and optimization criteria make ALM studies computationally very intensive task. It attracts attention to multiprocessor system and especially to the cheapest one – multi-core PCs and PC clusters.*

*In this article we are analyzing problem of parallel organization of portfolio optimization, results of using clusters for optimization and the most efficient cluster architecture for these kinds of tasks.*

## Introduction

Analysis of asset liability management (ALM) strategies especially for long term horizon is a crucial issue for banks, funds, insurance companies [1]. For example, pension funds in USA accumulate more than 6 trillion dollars and many of them are underfunded for 10-30%. Often ALM studies help to find investment/contribution strategies increasing the total return for 10% and are not costly.

The computer based ALM analysis requires:

- formal description DES of the investment and contribution strategies as a function with argument of economic state vectors for previous period of time and its value is a vector of characteristic of investment results (such as total return, risk and obligation estimations);

- formal description of economic possibilities or set of economic and liability scenarios SCEN – set of time series vectors of consistent indicators (it may be a set of historical data or data generated according some mathematical model);

- formal description of the objective functions - rules for the evaluation of outcome of the strategy.

These descriptions together we will call **Dynamic Decision Rule** (DDR).

Strategy can be estimated by apply DES for each scenario $s_1,\ldots, s_N$ of set SCEN and statistically analyzing of the results $y_1,\ldots, y_N$. This process can be represented with the following diagram:

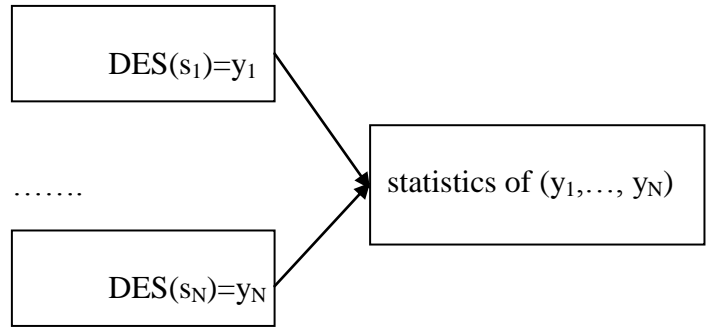

Figure 1.

Scenario generation based on mathematical model is a more popular way than using historical data collection just because it is difficult to say that some historical collection is full enough or frequency of the different types of historical scenarios well reflects the history.

Computational difficulties in ALM studies come from complexity of management strategy itself and number of scenarios for fully representation of all connections between economic and financial indicators. Our experience is that DDR may be equivalent up to 10 000 line of C++ code. Common number of indicators in the model is around 100 and each equation may depend on several other indicators. So we should talk about 1000-10000 scenarios. Time horizon may be from 10 to 320 periods.

But the problem becomes really hard when we want to find optimum parameters for an investment strategy, for instance, we want to find optimum portfolio allocation. In this case we have much more calculation intensive diagram:

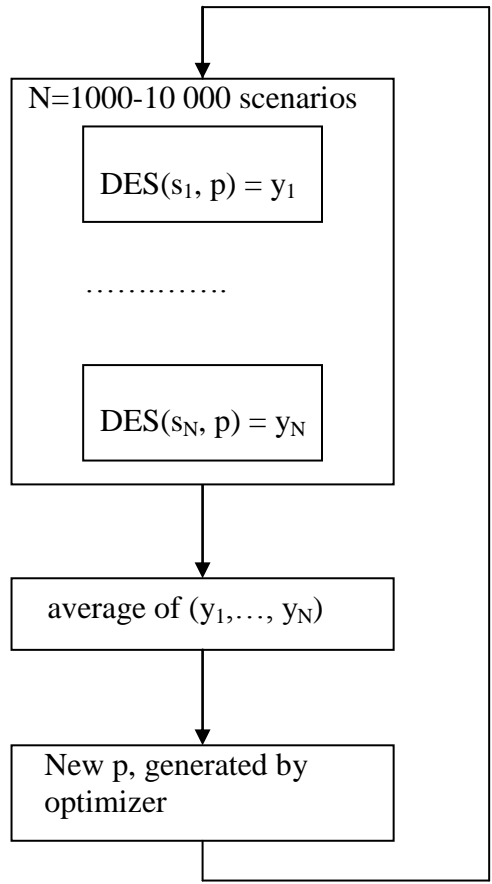

Figure 2.

Parameters of portfolio allocations may be tens and each of them is represented with hundreds points.

In ordinary practice, objective function and parameter domain are not convex and only heuristic methods can be really applied. Quality of the result of these methods depends on how many steps it is allowed to make. So number of possible steps here might be another 10 in degree tens.

All this shows that investment strategy optimization is a massive calculation task and to find a way for using multiple processors machines is an important task.

Parallel execution of schema above can be done in two ways. First is to find optimum for full scenario set but parts of parameter domain independently in each node and then select extreme value among the nodes. Second way is to find optimum for each parameter and part of scenario set independently in nodes and then select extreme values and repeat all for another parameter. For the first way we have to distribute parameter domain parts among nodes with full scenario set each and letting nodes to process parameters with a full set of scenarios. For the second way we have to distribute parts of scenario set and process them in parallel with each parameter point.

Plus of the first way is a low rate of communications, minus are duplication a scenario set and dividing a parameter domain to equal parts which might be a hard problem itself. Pluse of the second way is that no needs to distribute the scenario set, minus is that communications are needed for each testing parameter.

Further we will consider a second way of parallel calculation and avoid a problem of parameter set equal distribution.

There are several studies [1, 2] for using multiprocessor machines but usual result is that acceleration ratio rapidly goes down with growing number of processors. Another problem is that ALM studies are based on actuarial data and companies do not like to process them out of their own machines. Solution may be in using multi-core PCs or/and cluster of PCs connected via network.

Here we describe a schema for parallel optimization on the cluster of PCs and give a result of acceleration with a better behavior: it linearly increases with number of scenarios and transaction expenses grow logarithmically with a number of machines. The results are interesting for multi-core PCs as well because a multi-threading is resource consuming software too, and it adds a delay for total execution time. Real improvement would be if processors be able to communicate without involving operating system like special processors work.

## Cluster architecture

The calculation schema, shown in Figure 2, has two important features.
First, calculations DES for different scenarios are independent – calculation order does not matter for final result. Second, statistic of $(y_1,…, y_N)$ is a simple calculation problem and also may be done efficiently in a parallel. We will discuss it after making decision about communication graph for cluster.

Process of optimization (like efficient Glover's Tabu search [2]) consists of calculations DES with whole set of scenarios, points of optimization parameter space and light analysis of its result. So success in acceleration of applying DES to the set of scenarios will be helpful for whole optimization.

We start studies of hardware architecture with cheapest one and increase it complexity only if it is necessary for efficiency.

Let's consider Master – Slaves architecture. It uses standard network cards and hub. Optimization is based a two level tree control. Master is on the root, has a whole task and shares with Slaves some subtasks and data.

First question for using cluster is how to distribute scenario set between it nodes. It is clear that the number of scenarios in nodes has to be close to the ratio **Number_Of_Scenarios/Cluster_Size**.

For example, if we have 100 machine and 10090 scenarios, this ratio is 100.9. The solution to put 100 scenarios to first 99 machines and rest 190 scenarios to put to $100^{th}$ machine far from good. It almost twice exceed the time for processing bigger number of 10100 scenarios when each machine processes 110 scenarios.

The following simple algorithm guarantees that numbers of scenarios in cluster machine may differentiate only for 1.

The algorithm is :
    Portion_Size = [Number_Of_Scenarios/Cluster_Size]  // [ ] means integer part
    Rest = Number_Of_Scenarios-Portion_Size*Cluster_Size
    For each machine I // loop across cluster size
    If ( I < Rest)
    Send_To_Machine Portion_Size+1 scenarios
    If (I>= rest)
        Send_To_Machine Portion_Size scenarios

So if we have 100 machine and 10090 scenarios, then the first 90 machine will process 101 scenarios, and the last 10 will process 100 scenarios.

Suppose that set of 6000 scenarios is distributed equally in the 6 machine system above (master will get its part of task as well). So each machine gets 1000 scenarios.

For optimization, the cluster runs the following loop (here 5000 optimization parameter points is a limit if optimum was not found early).

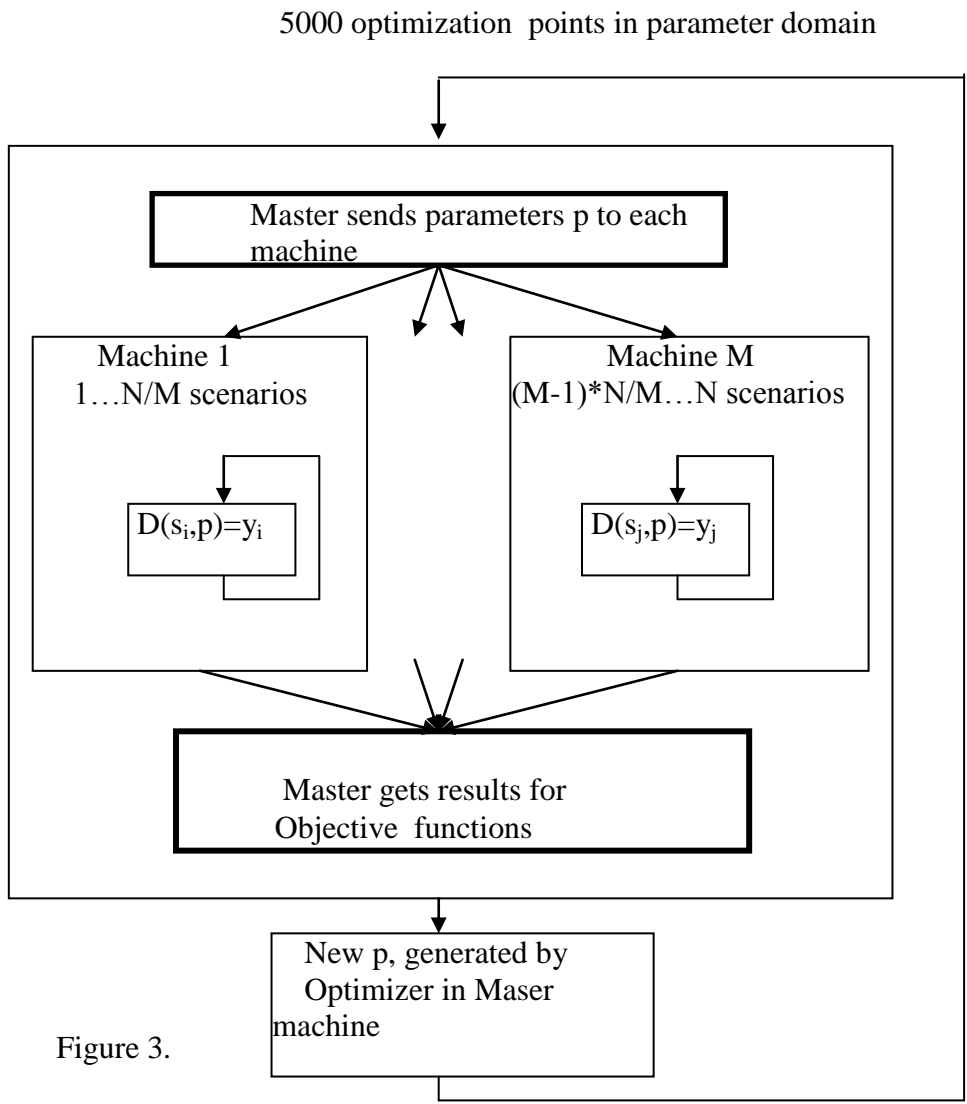

Figure 3.

Master selects a parameter vector using Tabu search algorithm. Then it sends that parameters to each Slave and all Slaves and Master are running DES on their piece of scenario sets and with the same optimization parameters. Extra operations here are sending small amount of parameters.

After processing its own piece of scenario set, Master collects the result from each Slave and calculates statistics (and objective function as a part of it).

The following is a time diagram for exchange with data package < rule, scenario Matrix>. In this diagram S means sending, G – getting data, W – waiting for call, R – evaluation

| Time | Master | Slave1 | Slave2 | Slave3 | Slave4 | Slave5 | Slave6 | Slave7 | Slave8 | Slave9 |
|------|--------|--------|--------|--------|--------|--------|--------|--------|--------|--------|
| 1 | S(Slave1) | **G** | W | W | W | W | W | W | W | W |
| 2 | S(Slave2) | W | **G** | W | W | W | W | W | W | W |
| 3 | S(Slave3) | W | W | **G** | W | W | W | W | W | W |
| 4 | S(Slave4) | W | W | W | **G** | W | W | W | W | W |
| 5 | S(Slave5) | W | W | W | W | **G** | W | W | W | W |
| 6 | S(Slave6) | W | W | W | W | W | **G** | W | W | W |
| 7 | S(Slave7) | W | W | W | W | W | W | **G** | W | W |
| 8 | S(Slave8) | W | W | W | W | W | W | W | **G** | W |
| 9 | S(Slave9) | W | W | W | W | W | W | W | W | **G** |
| 10 | R | R | R | R | R | R | R | R | R | R |
| …. | … | | | | | | | | | |
| 1000 | R | R | R | R | R | R | R | R | R | R |

Figure 4.

This schema has a bottleneck - access to the network bus is a sequential. For a big number of machines it may cause a growing delay.

In this case it makes sense to use more complex tree architecture as shown in next diagram.

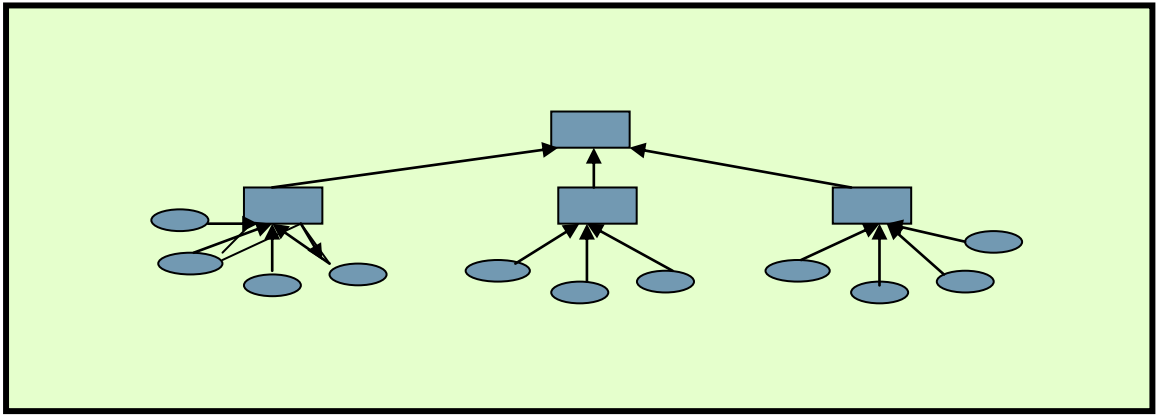

Figure 5.

Here we have two level Masters. First level master sends DDR and scenario data to second level masters and running the same loop as above. Second level masters are doing the same – sends task and data to its slaves but at the end of the loop's body they are collecting data from their slaves and sending result to upper level master. Second level masters are sending result to the upper level and so on.

This schema avoids a communication bottleneck for one level schema and made cluster power scalable in a big range.

Following diagram shows the case with two slaves for each second level Master.

| Time | Master | MSlave1 | Slave2 | Slave3 | MSlave4 | Slave5 | Slave6 | MSlave7 | Slave8 | Slave9 |
|---|---|---|---|---|---|---|---|---|---|---|
| 1 | S(Slave0) | **G** | W | W | **G** | W | W | **G** | W | W |
| 2 | S(Slave1) | W | **G** | W | W | **G** | W | W | **G** | W |
| 3 | S(Slave2) | W | W | **G** | W | W | **G** | W | W | **G** |
| 10 | R | R | R | R | R | R | R | R | R | R |
| ... | ... | | | | | | | | | |
| 1000 | R | R | R | R | R | R | R | R | R | R |

Figure 6.

## Ring architecture

To use tree architecture we have to have several hubs, in diagram shown above we need 4 hubs. Ring architecture may have the same time diagram, but without hubs and that is why it cheaper.

Consider a simple ring:

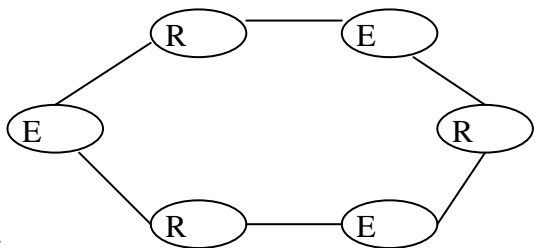

Figure 7.

Technically it may be created by putting two NICs in each PC and creating local networks for each neighboring pare.

Communication behavior for each PC is following.
E marked PCs sends data to the left immediate neighbors (marked R) and then gets data from the right immediate neighbors. Each R marked PC first gets data from its right neighbors marked E and then sends data to left neighbors E. So each even time communication diagram is this

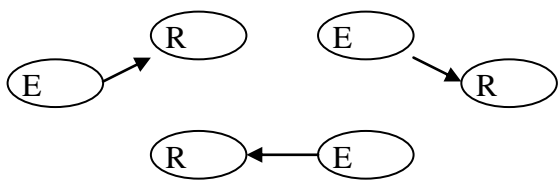

Figure 8.

And each odd time the communication diagram is this

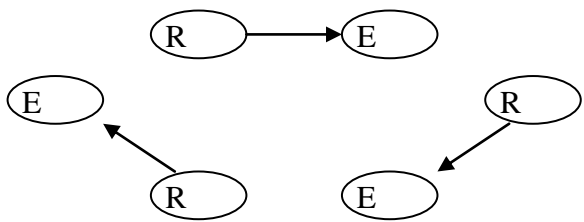

Figure 9.

Let's discuss the optimization calculation for statistic with two function: mean() and stdev() with arguments $(y_1,\ldots, y_N)$.  First result will be assigned to variable **a** and second, stdev value, to variable **b**.

Suppose, Q – is a cluster size, M=N/Q and machine k gets values $y_{Mk+1},\ldots y_{M(k+1)}$. To calculate mean$(y_1,\ldots y_N)$ we may calculate sum$(y_{M*k+1},\ldots y_{M*(k+1)})$ in machine k. In ring cluster each machine may send its sum and get result from the right neighbor.  It just adds the getting number to his result and send number getting from the right to his left neighbors. It is clear that after number of steps equal to cluster size each machine will have whole result.

For tree architecture case Slaves send their result to Master who will add all of partial sums, divide them by N and get mean value. Similarly is stdev calculation.  Each Slave may calculate sum of  $y_{Mk+1},\ldots y_{M(k+1)}$ and sum of their squares. And Master will just summarize them and get a square root according to the well known formula for calculation stdev without calculation its mean.

If the number of scenarios is much bigger than the number of Slaves, the biggest part of calculation will be done in parallel.

Communication is more efficient in a ring of rings architecture when machines in a ring (lets call it top level ring) are replaced with another lower level one and so on. Generally the communication graph looks like at the next picture. Each oval represents a separate network, yellow small rounds are machines in a ring.

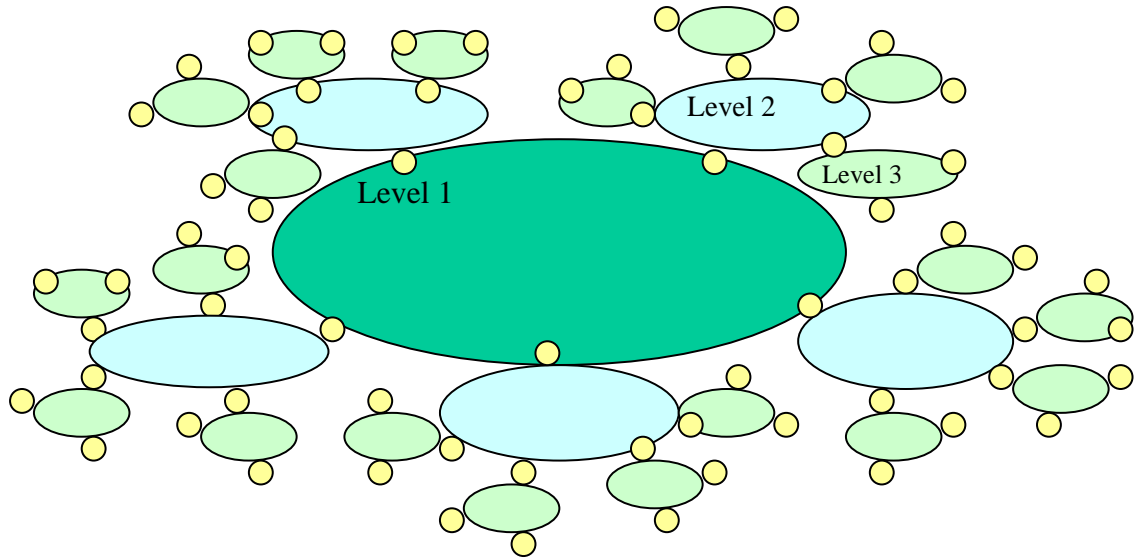

Figure 10.

One of the most attractive features of the architecture is that local degree of the graph does not exceed 4. So numbers of network interface cards (NICs) never exceed 4 in each machine.

Communication logic and technical devices for ring architecture cluster is a bit more complex than in tree. In case like scenario set processing when communication traffic is weak it is better to use more simpler tree architecture we considered first. Difference of the execution time between ring and multilevel tree architectures should not be really big. Another advantage of rings architecture is better reliability and isolation of broken elements but this is out of our issue here.

## Smooth objective function in a high dimension space

Another calculation intensive part of the application is a finding min for objective function with gradient method.

Here we need to calculate objective function F and its gradient $df/dx_1,\ldots df/dx_m$ in a large number of points while doing come down.

Let us consider case when optimization area is a hyper plain with linear bounds in n – dimensional space.

Let's $Y = (y_1, \ldots, y_m)$ is an independent basis and a matrix A with dimension (n-m, m) is a linear combination for other variables.

We want to distribute matrix A on cluster so that communication traffic during calculations of the Objective function and its gradient would be minimal without duplication of steps in cluster machines. And again our goal is to find big pieces of calculation which can be done in parallel and that is why ratio "additional operations for parallelizing"/ "sequence calculation" will be close to 0.

Here the main part of Objective function calculation is a calculation of dependent variable values. The value is a scalar product of some row to independent variable vector. If matrix rows are distributed on cluster machines then calculation may be done without communications. Also it is good that we need to distribute Matrix only once for the whole optimization which may runs hours. Master will get the result value from each machine and will finish the Objective function calculation.

Whole time consists of time $T_I$ for sending vector of independent variables to each machine, time $T_p$ for calculating one portion of dependent variables and $T_D$ for sending the result vector to Master, $T_p = m*T_c/$ Cluster Size where $T_c$ is a time for scalar product calculation.

So we have sequence calculation time $m*T_c$ and $(m/Cluster\_Size)*T_c+T_I + T_D$ for parallel one. Superiority of a parallel calculation of objective function depends on m: $m*(1-1/Cluster\ Size)*T_c>T_I+T_D$.

For gradient calculation one machine have to have some rows and some columns of the matrix A. Indeed,
$df/dx_i = SUM( df(y_1,y_2,..,y_n)/dy_j)*(dy_j/dx_i))$,
And $dy_j/dx_i = a_{ji}$.
To calculate df/dx we generally need to calculate f and that is why we need matrix rows and to calculate $dy_j/dx_i$ we need matrix columns. In addition to distributing rows we need to distribute columns. There is a cross of column and rows but in most case it make sense to have additional memory usage (which is $m*(m-n)/(Cluster\_Size)*2$) than to send back and force this amount of data for each gradient calculation. And number of gradient calculations may be big even for simple Objective function.

Because all $(m-n)*m$ elements of the matrix are generally involved in the calculation its computational complexity cannot be less than that.

**Remark.** Of course, there are a lot of exceptions. For example, if matrix contains big number of equal elements, it size can be shrunk and number of operation will be lowed.

But we are not using these sorts of acceleration and we want to distribute this matrix in cluster machines. Because we need to calculate depending on coordinate values in order to calculate Objective function, we will distribute rows among the cluster machines. Then depending variables calculation will be done in a parallel and after that result m-dimension vector will be send to master. Effectiveness ratio here is $O(1/n-m)$, operation weight in each machine is $O(m)$.

# Numerical experiments for Tabu optimization

In portfolio studies the most time consuming task is an optimization. This task has two imbedded loops: Tabu search loop (in all cases here it is 6000 iterations) and inside loop across scenarios for objective function calculations (see figure 3).
In our parallel calculation we will run in parallel the inner scenario loop. Master will run Tabu search iteration and define some values for optimization parameters, then it will use slave machines for Objective function calculation with these parameters. At the beginning it divides the set of scenarios into equal portions for each machine and send them to each slave together with a program for Objective function calculation. So if all machines are equally powerful the running time will be approximately the same. After finishing his part of calculation Master will gather information from Slaves and get average of them as an Objective function value.

Next table shows time of computation for different scenario set sizes and different cluster size. Clusters consist from 2.4Gh / 1GB Intel machines with Windows Server OS and communication is based on 1Gb network. ALM software is written in C#, communication part is based on **TcpListener** and **TcpClient** classes in one case and UdpClient at the another case. Time scale is a minutes. We selected a heavy ALM optimization for international company operating in 7 world markets so the time differences is less sensitive to the differences in hardware and easer to see.

| # | scenario number | number of machines | Duration Tcp | transaction time Tcp | Duration Udp | transaction time Udp |
|---|---|---|---|---|---|---|
| 1 | 500 | 1 | 27 | | | |
| 2 | 1000 | 1 | 52 | | | |
| 3 | 1500 | 1 | 79 | | | |
| 4 | 2000 | 1 | 98 | | | |
| 5 | 2000 | 2 | 133 | 81 | 59 | 7 |
| 6 | 3000 | 1 | 152 | | | |
| 7 | 3000 | 2 | 159 | 80 | 79 | 7 |
| 8 | 4000 | 1 | 266 | | | |
| 9 | 4000 | 2 | 178 | 80 | 105 | 7 |
| 10 | 6000 | 1 | 310 | | | |
| 11 | 6000 | 2 | 232 | 80 | 159 | 7 |
| 12 | 6000 | 3 | 250 | 152(2 comm) | 140 | 14(2 comm) |
| 13 | 10000 | 3 | 324 | 156 (2 comm) | 183 | 15(2 comm) |

Figure 11.

Last two columns show transaction times for Tcp and Udp cases estimated for cluster size 2 and 3.

Let **m** be a number of slaves, $T^c_{scen}$ be a cluster time for given number of scenarios **scen**, $T^s_{scen}$ be a time for processing number of scenarios **scen** on single machine. Then there is a formula for approximation cluster time
$$T^c_{scen} = T^s_{scen/m} + m * \text{TransactionTime},$$
or

$$\text{TransactionTime} = (T^c_{scen} - T^s_{scen/m})/m.$$

We may evaluate error for this formula using Table on Figure 11. It has columns for a transaction times calculated with this formula and fluctuation did not exceed 7%.

For line 5 number of slaves is 1 and number of scenarios is 2000, time for optimization with 1000 scenarios on stand alone machine is 52 (line 3) and
**Tcp TransactionTime** = 133-52 =80, **Udp TransactionTime** = 59-52 = 7.
For 10,000 scenarios and 3 machines we have $T^c_{3333} = T^s_{3000} * 33333/3000 = 152 * 1.1111 = 169$.
Here we are using dependency (and the table conforms it) $T^s_{scen/m} = T^s_{scen}/m.$

**Tcp TransactionTime** = 324-169 for a machines and 77.5 for each machine, for **Udp** protocol **TransactionTime** =15 or 7.5 for each machine.

So transaction time is really big and that is why running 6000 scenarios on 2 machine cluster with Tcp/IP protocol is faster than running the same scenarios on 3 machine cluster (lines 11,12). Another fact is that cluster works faster only starting from big number of scenarios – more than 3000 for Tcp case and more than 150 for Udp case.

## Cluster topology

Remarkable amount of transaction time also makes a cluster topology to be an important issue. For one level architecture

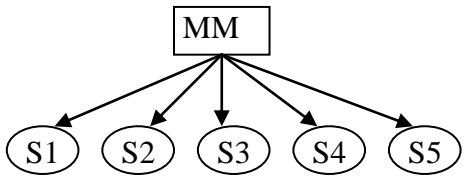

Figure 12.

transaction time is growing with number **m** of slaves linearly
$T^c_{scen} = T^s_{scen/m} + m * \text{TransactionTime}.$

We may estimate minimum number of machines when adding new machine does not decrease the time. It happense for the min **m** when
$T^c_{scen} / \text{TransactionTime} < m * (m-1),$

For **scen**=6000, **TransactionTime** = 75 (line 12) smallest **m** for Tcp case is 3: 250/75<3*2, for Udp **m**>7 (250/7<7*6).
For 10 times more number of scenarios **m**>6 and **m**>18 respectively.

But this is correct only for one level tree.

For the two level architecture this time may be less.

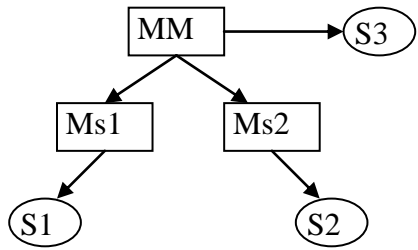

Figure 13.

For example, collecting result to main master MM goes in this order: simultaneously S1 sends data to Ms1, S2 sends its data to Ms2 and S3 sends data to MM. Then Ms1 sends data to MM and then Ms2 sends data to MM. Transaction time in first case is **5*TransactionTime**, in second – **3*TransactionTime**.

Important issue here is an order for transaction services. Also it is clear that optimum architecture is not a binary tree with node's local degree less than 3 but a tree created with the following recursion procedure:

**Start:** add node and labeled it 0.
**Each next step n:** add edge and node to each node added on previous steps and label this node with this step number.

On next picture is shown 3 steps of this procedure. Each stage of creation is in rectangular, numbers in circles are a step number when it was added.

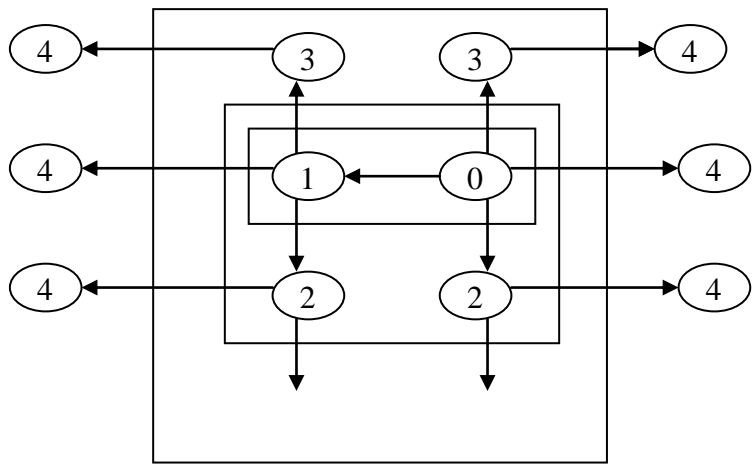

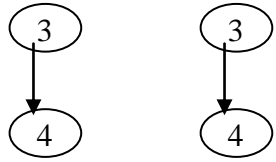

Figure 14.

Let $N_i$ be a number of nodes added at time i, $S_i$ all nodes added at time 1, 2, …, i. Then $N_{i+1}=S_i$, $S_i=\text{sum}(N_j, j=1,..,i)$ and hence $N_{i+1}=N_i+\text{sum}(N_j,j=1,…,i)$ or $N_{i+1}=N_i*2=2^k*N_{i-k}=2^i$.

So for any step **n** the procedure is
- adding $2^n$ nodes
- number of nodes in a whole graph is $2^{n+1}$.

For any number of nodes P optimal connection graph is a graph build on
k =integer($\log_2 P$) steps and added P-k nodes as a leaves to any of nodes from $2^k$ optimal graph.

Transactions proceed according the labels.
Sending info from main master to all cluster members goes in the following order. Each machine already getting info sends it to his slave labeled with minimum number among not served yet. According to the creation procedure, for each node its nearest node will have different labels. That is why one node will be involved only in one transaction.

Optimality of this topology comes from the fact that at any period of time each machine who got the info is involved in sending info to someone without the info. The number of sequential transactions is a next integer to **$\log_2 m$, m** is number of cluster machines.

For 16 machine optimal cluster transaction time will be **4*TransactionTime**, for one level cluster this number will be **15*TransactionTime**.

Running time for optimization on optimized cluster is defined with formula
**TranssactionTime*$\log_2 m$+$T^s_{scen/m}$**. For 4 machine cluster acceleration for 10000 scenarios will be 1.32 and for 6000 - 1.47 .

**Efficiency graphs**

On Figure 15 are shown graphs of dependency of execution time and cluster size for 1-level tree and optimal tree.

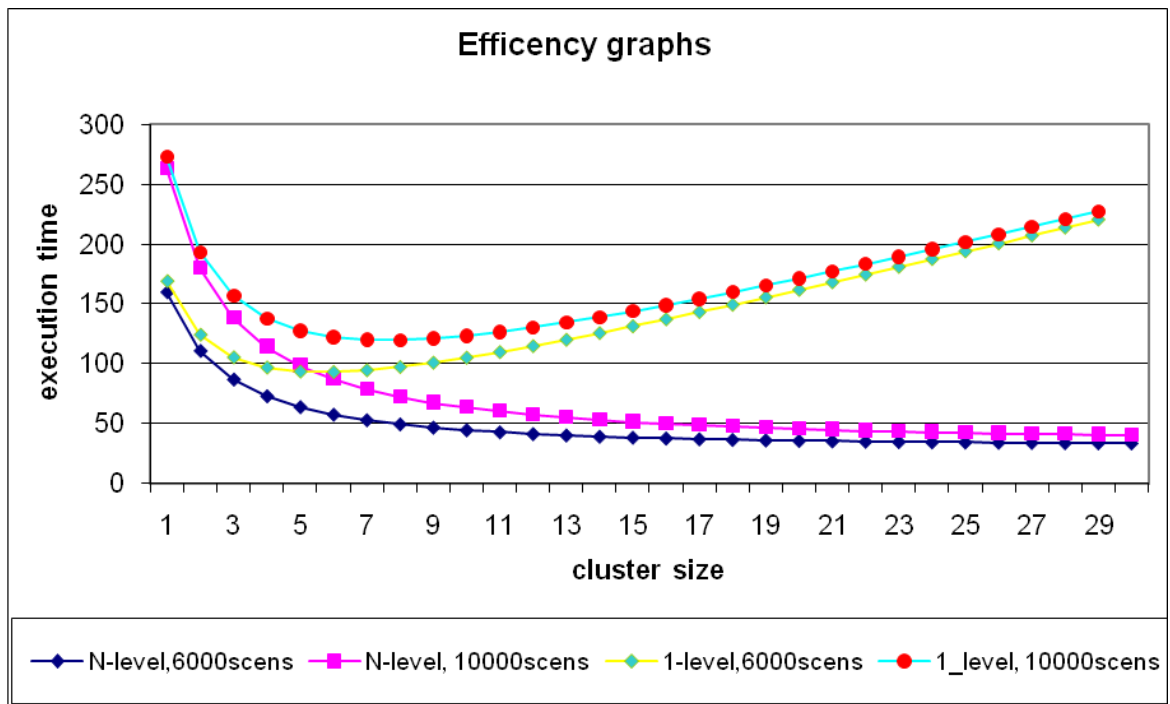

Figure 15.

For 1-level and 6000 scenarios tree if number of machines exceed 6 then adding new machines increase the time. For 1-level and 10000 scenarios this number is 8.

For optimal tree the adding new machine to the cluster decrease time for any number of machines (of cause for less than scenario number). But percentage of efficiency of adding new machine goes down starting from 10.

Usually one ALM study requires 30-100 optimizations, so gain will be days of professional economists time for each study. Big companies are doing hundreds of studies per year and such gain is a remarkable.

# Conclusion

Using real cluster experiments we have shown that for ALM type of tasks with highly parallelism and low data traffic architecture still crucial issue: increasing transaction speed to 10 times improves total time only twice. We found cluster architecture and communication algorithm for this type of tasks where processing time is decrease linearly and transaction time growth only logarithmically.